\documentclass[conference]{IEEEtran}
\IEEEoverridecommandlockouts
\usepackage{cite}
\usepackage[utf8]{inputenc} 
\usepackage[T1]{fontenc}    
\usepackage{hyperref}       
\usepackage{url}            
\usepackage{booktabs}       
\usepackage{amsfonts}       
\usepackage{nicefrac}       
\usepackage{microtype}      
\usepackage{amsmath}
\usepackage{amssymb}
\usepackage{graphicx}
\usepackage{mdframed}
\usepackage{listings}
\usepackage{wrapfig}
\usepackage{float}
\usepackage{xcolor}
\usepackage{enumitem}
\usepackage{graphicx}
\usepackage{caption}
\usepackage{verbatim}
\usepackage{subcaption}
\usepackage{authblk}
\usepackage{multirow}

\def\BibTeX{{\rm B\kern-.05em{\sc i\kern-.025em b}\kern-.08em
    T\kern-.1667em\lower.7ex\hbox{E}\kern-.125emX}}

\begin{document}

\title{Graph learning methods to extract empathy supporting regions in a naturalistic stimuli fMRI}

\author{Sasanka GRS$^{\star}$ \qquad Ayushi Agrawal$^{\dagger}$ \qquad Santosh Nannuru$^{\star}$ \qquad Kavita Vemuri$^{\dagger}$ \\
$^{\star}$ Signal Processing and Communication Research Center, IIIT Hyderabad, India \\
$^{\dagger}$ Cognitive Science Lab, IIIT Hyderabad, India}

\maketitle

\begin{abstract}
Functional MRI (fMRI) research using naturalistic stimuli like movies, examines brain network interactions supporting complex cognitive processes like empathy. Multiple brain areas like Insula, PFC, ACC and parietal constitute the empathy network. Applying graph learning methods to whole-brain timeseries signals, we propose a novel processing pipeline that includes high-pass filtering, voxel-level clustering, and windowed graph learning with a sparsity-based approach. The study involves two short-movies shown to 14 healthy volunteers. A total of 54 regions extracted from the AAL Atlas were considered for the study. The sparsity-based graph learning method consistently outperforms others in capturing variations in the emotion contagion scale, achieving over 88\% accuracy averaged across participants. Temporal analysis reveals a gradual induction of empathy, supported by the method's effectiveness in capturing dynamic connectomes through graph clustering. Edge-weight dynamics analysis underscores the superiority of sparsity-based learning, with others either hardly activating or giving noisy activations. Connectome-network analysis highlights the pivotal role of the Insula, Amygdala, and Thalamus in empathy, with lateral brain connections facilitating synchronized responses. Spectral filtering analysis emphasizes the significance of the band-pass filter in isolating regions linked to emotional and empathetic processing during empathy HIGH states. Key regions like Amygdala, Insula, and Angular Gyrus consistently activate, supporting their critical role in immediate emotional responses. Strong similarities across movies in graph cluster labels, connectome-network analysis, and spectral filtering-based analyses reveal robust neural correlates of empathy. These findings enhance the understanding of empathy-related neural dynamics and lay ground by identifying regions specific to empathetic response to stimulus, paving way for targeted interventions and treatments for conditions associated with empathetic processing.
\end{abstract}

\section{Introduction}
\label{sec:intro}

Empathy is our ability to take the perspective and share the emotions and feelings of others. The process includes mentalizing or cognitive evaluation and emotional response, making it a complex construct to study from behaviour and brain activity. Studies have looked at empathy response in conditions like Autism spectrum \cite{Ruther}, psychopathy \cite{Decety}, age-related \cite{Beadle} and social interactions \cite{Akitsuki}. In the recent years, the role of context or situation setting in differential empathy response has been recorded \cite{response}. First, naturalistic stimuli like movies, or text narratives present an ecologically valid context to participants in an fMRI or EEG experiment, as real-life experiences are complex. Second, such stimuli can provide insights into brain networks, processing complex sensory information. Using techniques like fMRI, experiments have revealed that the following constant but distributed brain areas \cite{Singer,Fan,Jackson,Ruby,Lamm}, like the Anterior Cingulate Cortex (ACC) (attributed to decision-making, social cognition, judgements and emotion regulation), Insula (emotion, self-awareness, introspection and critical for pain empathy), the Inferior frontal gyrus which supports a host of functions from language, memory, and cognitive driven judgements on other’s emotional states, Amygdala (emotion processing especially fear), and the Superior temporal gyrus, Somatosensory cortex, the areas forming the default mode network (a resting state network).

Movies (short or full length) as stimuli in fMRI provide the context and the complex sensory input, which allows the viewer to form a scene-dependent empathy response, for analysis of the temporal dynamics of the brain regions \cite{Hasson,VemuriA,Tikka}. Long-duration stimuli are a challenge to analyse using the traditional general linear model. Typically, general linear models are used to separate the signal from noise and time-delays introduced at the experiment design stage. To study whole-brain dynamics for naturalistic stimuli, statistical methods like Independent Component Analysis \cite{VemuriB}, Multivoxel Pattern Analysis \cite{Zaki} and Inter-Subject Correlation \cite{Kauppi} have been extensively applied. In recent times, graph learning and signal processing have been efficient in extracting functional connectivity networks \cite{Chen,Fau}, which show quasi-stable states and individual differences. The questions we endeavour to answer by this preliminary study are: given the complexity of empathy brain networks and the multi-modal stimulus used, can we build methods to model whole brain dynamics reflecting the behavioral responses to the narrative? This paper explores the application of various graph learning techniques to fMRI data for extracting region-level activation, analysing the dynamic functional connectivity of empathy-supporting brain regions within a whole-brain setting, and  validating the findings by comparing emotional scale ratings with the graph clusters plotted over time.


Graph Signal Processing (GSP) \cite{GSPBasics} is a versatile field that merges graph theory, linear algebra, and signal processing to analyze signals on graphs, offering a robust framework for extracting insights from networked data. In brain signal analysis, GSP has been applied for tasks such as filtering brain activity using graph spectral modes \cite{Huang} and signal decompositions \cite{Huang1}. \cite{Huang1} and \cite{pearson} use distance-based and Pearson's correlation-based approaches to functional brain connectivity. In the realm of studying brain connectivity through graph learning, previous methods \cite{Gao} relying on smoothness assumptions are outperformed by our sparsity-based approach. In comparison to existing methods like \cite{Sparsity1}, our proposed pipeline maintains subject-specificity without employing inter-subject group sparsity regularization. Unlike \cite{Sparsity2}, which assumes temporal coherence with sparse corruptions for disease classification, and \cite{Sparsity3}, which assumes spectral sparsity, our approach effectively applies sparsity-based constraints to task-based data, showcasing versatility beyond resting state scenarios. Contrasting with prior research employing Graph Neural Networks (GNN) \cite{GNN1}, Graph Convolutional Networks (GCN) \cite{GCN1} for task-based problems, our work distinguishes itself by focusing on data-driven graph learning. We extract meaningful graph connectivity patterns directly from the data using optimization methods, setting our approach apart from studies primarily using GSP for task-oriented applications.

In our study, graph nodes represent brain regions, each node containing a time-series fMRI BOLD (Blood Oxygen Level Dependent) signal, further processed using the proposed pipeline before being used for learning the functional connectivity. We aim to learn the underlying brain functional connectivity among these regions from collective fMRI activation signals, particularly focusing on long-duration movie stimuli (it is a short movie, but from an fMRI standpoint, it is a long-duration movie). The learned graphs' edge-weights quantify connectivity strength between different brain regions, allowing us to examine dynamic changes in edge-weights among different brain regions over time and variations in fMRI signals across the emotion scale. These graphs act as intermediate results, laying the groundwork for subsequent analyses and the derivation of more detailed outcomes.

Utilizing Graph Signal Processing (GSP) and graph learning is motivated by the brain's intricate functional organization. Unlike traditional region-specific analyses, GSP allows a nuanced exploration of connections between different regions, capturing the brain's dynamic network. In empathy tasks, where multiple regions contribute, GSP captures complex network dynamics. It unveils patterns governing coordinated activity, fitting the brain's interconnected nature. Graph Fourier Transform (GFT) combines regional activations and connectivity graphs. GFT's spectral bands serve distinct purposes: the low-pass band emphasizes stable nodes, the high-pass band explores deeper localization, while the band-pass band strikes a balance, identifying patches without extremes, making it interesting for empathy-related brain activity exploration. Applying GFT aligns with our objective to understand empathy-specific regions efficiently.

This research presents novel contributions aimed at advancing the understanding of the neural correlates of empathy through fMRI data, which include -

\begin{itemize}
    \item Application of sparsity-based graph learning to fMRI data.
    \item Proposed a unique signal processing pipeline incorporating high-pass filtering and phase-based voxel clustering.
    \item Exploration of dynamic functional connectivity patterns and graph metrics in contrast to traditional regional activation levels.
    \item Analysis of raw fMRI BOLD signals without statistical component-based analyses for extracting dynamic functional connectivity.
    \item Implementation of Graph Fourier Transform for isolating task-specific groups of regions using spectral-domain band-pass filtering.
\end{itemize}

The paper is outlined as follows, in Section \ref{sec:data}, we give a detailed description of the data collection and pre-processing. In Section \ref{sec:preliminaries}, we introduce notations and pre-requisites for the study. In Section \ref{sec:methodology}, we discuss our methodology which processes raw BOLD signals to obtain the functional connectivity networks, processed in time windows. Subsequently, we discuss the analyses, motivation behind them and a detailed procedure to be performed over the obtained graphs in Section \ref{sec:analyses}. Further, we present the results in Section \ref{sec:results} across different time instants and using various methods. Analyses are also performed for specific brain regions and connections which are known to play an important role in an empathetic response. Finally, we discuss our reported results, and back the claims in Section \ref{sec:discussion}, followed with a conclusion of the work, its limitations and future directions in Section \ref{sec:conclusion}.

We analysed the results using MATLAB and Python on an NVIDIA RTX 3050 GPU. While we cannot provide open access to the data now, we are open to sharing upon request. The codes are available at \href{https://github.com/Sasanka-GRS/Brain-Connectivity-fMRI}{https://github.com/Sasanka-GRS/Brain-Connectivity-fMRI}.

\section{Data acquisition and pre-processing}
\label{sec:data}

Processing raw BOLD signals is a crucial step in fMRI data analysis, as it helps to remove noise and artifacts and extract meaningful information from the data. Pre-processing steps, such as slice-timing correction, motion correction, normalisation, and spatial smoothing, are typically applied to the raw fMRI data to improve the quality of the data and reduce sources of noise.

\subsection{Participants and approvals}

Sixteen healthy subjects (9 male and 7 female) participated in this study. For the preliminary graph learning method application reported in this paper, 14 participants' data were considered. In addition, 40 participants with similar backgrounds participated in a survey to rate the movies. 

The consent form mentioned that two short movie clips of 8 minutes 45s will be shown in the scanner. Additionally, as a part of the consent form and pre-scanning instruction, the experimenter informed the participant that they could quit the experiment at any time without any penalty. The human ethics committee of the Institute (IRB) approved the study, and all subjects provided a written informed consent. They received Rs.1,500 (\$18) in compensation for their time.

\subsection{Stimuli}

There were two short feature films used for the analyses: An Egyptian short film titled “These Times” (M1) and a Czech live action short film titled "Most", re-titled "The Bridge" (M2). The storyline for both the movies is provided in Annexure A. The movies were edited to run for 8 minutes 45s approximately. The rationale for considering the non-local language movies was - a) The movies had actors who did not have same ethnicity as the participants, to control for intra-cultural biases arising from economic status, caste, language, and physical characteristics and b) the direction was concise and had very poignant instances of change in narrative.

\subsection{MRI acquisition}

Data was collected from a 3 Tesla Philips fMRI 8-channel Achieva head coil whole-body scanner using gradient echo-planar imaging (EPI) sequence. Functional images were acquired with a repetition time (TR) of 2s, echo time (TE) of 35ms, and a flip angle of $90^{o}$ using a weighted echo-planar sequence. Other parameters include acquisition matrix of 64 $\times$ 64, slice thickness of 5mm, a gap of 1mm, 30 axial slices in the AC-PC plane, REC voxel MPS of 1.8 $\times$ 1.8 $\times$ 5.0 mm, and acquisition voxel MPS of 3.5 $\times$ 3.5 $\times$ 5.0 mm. The stimuli was designed in the E-prime software and projected from outside the scanner room onto a mirror mounted over the head coil. The total scan duration was 25 minutes, with movie clip analysed for this study shown for 8 minutes 45s. For the 3-dimensional T1-weighted structural data the Fast Field Echo (FFE) technique was used with a TR/TE = 8.39ms/3.7ms acquiring 150 slices, flip angle of $8^{o}$, field of view (FOV) : 250 $\times$ 230 and voxel volume of 0.98 $\times$ 0.98 $\times$ 1.0 mm. The functional scan acquisition and events in the paradigm were trigger-locked. 

\subsection{SPM-12 pre-processing of fMRI data}

The fMRI data were analyzed with SPM12 (Welcome Department of Imaging Neuroscience, London, UK) toolbox running on MATLAB R2019. Spatial pre-processing steps include: realignment (realigned to the mean image), coregistration (to each participant's individual T1 structural scan), segmentation (segmentation into grey, white matter and cerebrospinal fluid tissues), normalization (normalized to the Indian Brain Templates). The slice time correction was avoided as the TR was 2 seconds \cite{Poldrack}. The head movement correction was applied with the maximum head motion of < 2 mm and < $0.5^{o}$. The total number of scans was 262.

\subsection{Atlas and extraction of regions}

The selection of brain regions was conducted using the AAL Atlas, encompassing a total of 54 bilateral regions (27 on each side of the brain). The chosen regions were strategically picked to ensure comprehensive coverage, particularly focusing on frontal areas. Emphasis was given to empathy-specific regions such as the Insula, ACC, and Triangularis. In addition to these targeted regions, a broader set was considered to cover the entire brain while avoiding redundancy. It's worth noting that all 90 regions from the AAL Atlas were not included in the study, as the chosen regions were thoughtfully curated to provide ample representation across the frontal, parietal, occipital, and temporal lobes, ensuring a well-rounded coverage of the entire brain. This selection not only serves the purpose of studying empathy-associated regions but also optimizes computational efficiency by avoiding unnecessary redundancy, thereby expanding the practical applicability of the analysis. The list of regions chosen for the analysis has been provided in Annexure B.

The selected regions are obtained as masks from the AAL3 Atlas using the SPM12 toolbox in MATLAB. These extracted masks are then resized to match the data dimensions using the Nilearn and Nibabel libraries in Python. The original mask dimensions are $91\times 109\times 91$, while the data dimensions are $79\times 96\times 32$. Subsequently, the scaled masks are applied to the data, facilitating the extraction of time-series for all voxels within each region. These region-specific time-series are consolidated into 2D matrices. This process is repeated for all 54 regions, and the resulting matrices are stored in MATLAB struct files for subsequent analysis.

\section{Preliminaries}
\label{sec:preliminaries}

\subsection{Notations}

A graph $\mathcal{G}$ consists of a set of $N$ nodes $\mathcal{N} = \{1,2,\ldots,N\}$ and a set of edges $\mathcal{E}$. The graph connectivity is described using the adjacency matrix $\mathbf{A}$ where the $(i,j)$-th entry denotes the weight associated with the edge connecting nodes $i$ and $j$. The graph Laplacian matrix $\mathbf{L}$ is given by $\mathbf{L} = \mathbf{D} - \mathbf{A}$ where $\mathbf{D}$ is a diagonal matrix with the diagonal entries as the degree of each node. A graph signal $\mathbf{x} \in \mathbb{R}^N$ is a vector with the entry $\mathbf{x}[n]$ denoting the signal at node $n$ of the graph. For a $P$ dimensional signal at each node, the graph signal is an $N \times P$ matrix $\mathbf{X} = \left[\mathbf{x}_{1}, \mathbf{x}_{2} \hdots \mathbf{x}_{P} \right]$. $\mathbf{X}[n]$ denotes all the $P$ observations of the graph signal at node $n$ and $\mathbf{X}_{p}[n]$ denotes the $p$th observation of the signal at node $n$.

An important property of the graph Laplacian arises through its eigendecomposition. Mathematically, the eigendecomposition of the graph Laplacian is given as $\mathbf{L} = \mathbf{U} \mathbf{\Lambda} \mathbf{U}^{T}$ where, $\mathbf{\Lambda}$ is the set of eigenvalues, while $\mathbf{U}$ is the set of eigenvectors for the corresponding eigenvalues. The eigenvalue matrix $\mathbf{\Lambda}$ is diagonal matrix with the diagonal entries consisting of the eigenvalues $\lambda_{l},\ l=0,1,\hdots,N-1$ in the increasing order, and $\lambda_0=0$. These are the Fourier modes, also referred to as the graph frequencies, which holds frequency information of the graph. The set of eigenvectors $\mathbf{U}$ consists of $\mathbf{u}_{l},\ l=0,1,\hdots,N-1$, which are the Fourier bases corresponding to the Fourier modes.

\subsection{Graph Fourier transform (GFT) and spectral filtering}
\label{subsec: GFT}

The graph Fourier transform (GFT) of a graph signal $\mathbf{x}$ is mathematically defined as $\mathbf{\hat{x}} = \mathbf{U}^{T} \mathbf{x}$. The GFT transforms the signal from the graph domain to the spectral domain, capturing the frequency content of signals, revealing essential information about the underlying structure and connectivity of the graph. By decomposing a signal into its graph Fourier components, one can identify patterns, clusters, or anomalies within complex networks. Similar to the classical Fourier transform, the graph Fourier coefficient $\mathbf{\hat{x}}[l]$ serves as an indicator of the energy of the signal $\mathbf{x}$ at the corresponding graph frequency $\lambda_{l}$. 

Just as in the classical Fourier domain, it is possible to design spectral domain filters that operate on specific graph frequencies. An ideal filter $f(\lambda)$ in spectral domain can be written as

\begin{equation}
    f(\lambda; \lambda_{c}, \lambda_{1}, \lambda_{2}) = 
    \begin{cases}
        1 & \text{if } \lambda \leq \lambda_{c} \text{ (low-pass)} \\
        1 & \text{if } \lambda_{1} \leq \lambda \leq \lambda_{2} \text{ (band-pass)} \\
        1 & \text{if } \lambda_{c} \leq \lambda \text{ (high-pass)} \\
        0 & \text{otherwise}
    \end{cases}
\end{equation}

Here, the parameters $\lambda_{c}$, $\lambda_{1}$ and $\lambda_{2}$ determine the characteristics of the filter (cutoff frequencies for low-pass and high-pass, and the band for band-pass).

The inverse operation from the spectral domain back to the graph domain is mathematically defined as $\mathbf{x} = \mathbf{U} \mathbf{\hat{x}}$. From the above definitions, a graph spectral filtering operator can be defined as

\begin{equation}
    \mathcal{H}_{\mathbf{f}} \mathbf{x} = \mathbf{U}\mathbf{f} \odot (\mathbf{U}^{T} \mathbf{x})  
\end{equation}

where $\mathbf{f}$ is the spectral domain filter, $\odot$ represents an element-wise product. This work explores the brain regions identified in each of these bands, by splitting the graph spectrum into 3 equal-sized bands, i.e., the cutoffs $\lambda_{c} = \lambda_{max/3}$ for the low-pass band, $\lambda_{1} = \lambda_{max/3}$ and $\lambda_{2} = \lambda_{2max/3}$ for the band-pass band and $\lambda_{c} = \lambda_{2max/3}$ for the high-pass band, where $\lambda_{max}$ is the maximum frequency in the graph ($\lambda_{max} = \lambda_{N-1}$).

The different graph spectral bands offer valuable insights into the underlying structure and dynamics of brain connectivity. Intuitively, the low-pass band targets stationary nodes and predominantly captures components of the signal that exhibit consistency across the entire graph, highlighting regions with relatively stable behavior. Conversely, the high-pass band delves into deeper localization of nodes, emphasizing dynamic changes and localized variations within the network. Of particular interest is the band-pass band, which focuses on identifying patches within the graph. Unlike the extremes of the low-pass and high-pass bands, the band-pass band strikes a balance, highlighting regions with significant variations while avoiding the extremes. This characteristic renders it particularly intriguing for further analysis in subsequent sections of the paper, as it offers a nuanced perspective on the connectivity patterns associated with empathy-related brain activity.

\subsection{Graph Clustering}
\label{subsec:graph clustering}

In this study, the objective of graph clustering is to reveal underlying temporal patterns within a set of graphs, each corresponding to a distinct instant in time. Through hierarchical clustering with ward linkage, this step aims to minimize intra-class variance by flattening the adjacency matrix of each graph and subsequently clustering these flattened matrices over the entire time range. This clustering approach enables the categorization of graphs based on their inherent similarities, allowing us to explore the temporal evolution of functional connectivity patterns.

The specific interest in this work focuses on binary classification, aiming to differentiate between empathy-related and non-empathy-related graphs. The hierarchical clustering process employed facilitates the identification of clusters associated with distinct behavioral states over time. To assess the efficacy of any graph learning algorithm in extracting functional connectivity, particularly concerning neural activations linked to empathy, the clusters are examined in relation to the emotion contagion scale which will be discussed in the next subsection. The emotion scale acts as a notion for ground truth to compare with and will be used similarly in further analyses.


\subsection{Cross-Correlation and empathy score}
\label{subsec:cross-corr & score}

To evaluate the empathy levels of each subject against the ground truth, represented by the emotion contagion scale provided by participants, cross-correlation serves as the chosen metric. Cross-correlation, a mathematical operation, assesses the similarity between two signals by examining their alignment across various time shifts. In signal processing, it quantifies the resemblance between the shapes of two signals. By computing correlation coefficients at different time lags, cross-correlation indicates the degree of matching between the signals at different time points. This metric ranges from -1 to 1, where a value of 1 indicates a perfect match, 0 denotes no correlation, and -1 signifies perfect anti-correlation. 

Applied to time-series data, cross-correlation facilitates the evaluation of temporal alignment and synchronization between signals, providing a quantitative measure of the percentage match between their shapes. This analytical approach proves particularly useful for comparing patterns, identifying similarities, and uncovering relationships within diverse datasets, rendering it suitable for quantifying the empathy level of each subject. We define empathy score of a subject as the percentage match of the temporal graph clustering labels with the emotion scale, which is calculated using the highest cross correlation match between the two signals.

\subsection{Functional connectivity and Pearson-correlation-coefficients}

Functional connectivity is defined as the temporal coincidence of spatially distant neurophysiological events. This concept is based on the statistical relationship between the measures of activity recorded in different brain regions. It suggests that two regions exhibit functional connectivity if there is a statistical association between their activities, regardless of their physical proximity. Functional connectivity is purely correlative, meaning that it reflects observational measures of how different brain areas' activities coincide over time. This correlative nature doesn't imply a causal relationship or a direct physical connection between the regions.

One widely used method to assess functional connectivity is through Pearson's correlation coefficient \cite{pearson}. This metric quantifies the linear relationship between the time series of two brain regions, providing a measure of how synchronized their activities are over time. This coefficient, denoted as $\rho(\mathbf{x}, \mathbf{y})$, measures the correlation between two signals, $\mathbf{x}$ and $\mathbf{y}$. The formula is given by:

\begin{equation}
\rho(\mathbf{x},\mathbf{y})
= \frac{1}{N-1} \sum_{i=1}^{N}\left(\frac{\mathbf{x}_{i}-\mu_{\mathbf{x}}}{\sigma_{\mathbf{x}}}\right)\left(\frac{\mathbf{y}_{i}-\mu_{\mathbf{y}}}{\sigma_{\mathbf{y}}}\right)
\end{equation}

Here, $\mathbf{x}_{i}$ represents the $i$th element of the signal $\mathbf{x}$, while $\mu_{\mathbf{x}}$ and $\sigma_{\mathbf{x}}$ denote the mean and variance of the signal, respectively. Similarly, $\mathbf{y}_{i}$, $\mu_{\mathbf{y}}$, and $\sigma_{\mathbf{y}}$ correspond to the $i$th element, mean, and variance of signal $\mathbf{y}$. This correlation coefficient is directly employed in the construction of the adjacency matrix by calculating these coefficients pairwise.

\begin{figure}[t]
  \centering
  \includegraphics[width=\linewidth]{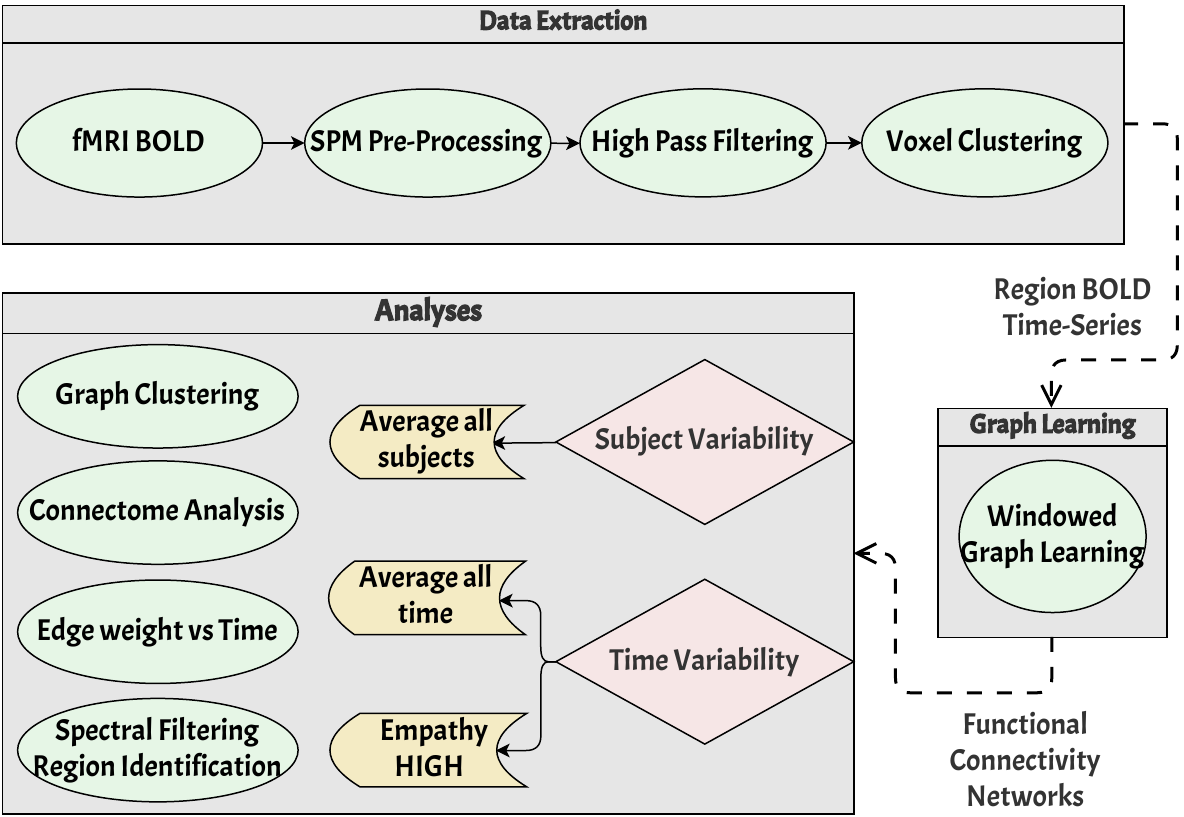}
  \caption{Pipeline used for fMRI BOLD signal functional connectivity analysis}
  \label{fig:pipeline}
\end{figure}

Although many different measures of functional connectivity exist, most human neuroimaging studies currently employ Pearson correlation of time-series as a widely adopted and well-established method in the field. This method is frequently applied in resting-state analyses, utilizing techniques such as time-series correlations in BOLD fMRI data acquired in a task-free state. In this study, we utilize this method as part of a comparative analysis to demonstrate the effectiveness of optimization-based graph learning techniques.

\section{Methodology}
\label{sec:methodology}

Figure \ref{fig:pipeline} shows the steps involved in extracting a time-series signal, at a voxel level and further at a region level, which can be passed into a graph learning module to extract time-windowed functional connectivity matrices. Detailed explanations of the various methods employed in the pipeline will be provided next.

\subsection{Normalization and High-pass filtering}

Initially, the raw BOLD signal undergoes normalization through mean-removal, followed by high-pass filtering of the time-series at each voxel. Normalization is a crucial step to ensure that simultaneous activations are accurately captured, particularly when the mean value of each region's time-series is biased and non-zero. By removing the mean, all signals are centered around 0 DC, enhancing the ability to capture small variations and temporal coincidences in the signals. This step proves vital, as demonstrated in the graph learning module outlined in Section \ref{subsec:GL}.

\begin{figure}
    \centering    
    \includegraphics[width=\linewidth]{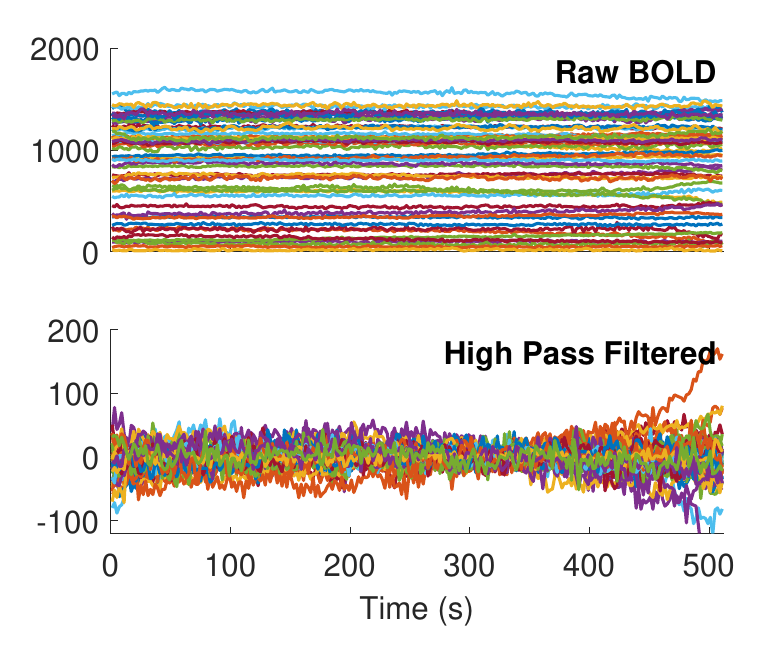}
    \caption{The raw BOLD and high-pass filtered BOLD signals at all the brain regions considered.}
    \label{fig:normalized}
\end{figure}

High-pass filtering is a critical pre-processing step in fMRI data analysis that removes slow temporal fluctuations in the signal that are not related to the neural activity of interest. These low-frequency components, including scanner drift, physiological noise and subject movement noise \cite{HPFNoise}, can mask neural activity and reduce the sensitivity of the analysis to the effects of interest. By removing these low-frequency components using high-pass filtering, the quality of fMRI data can be improved. The removal of slow-moving signals from fMRI data can be particularly important in studies investigating the dynamics of brain networks \cite{HPFDynamics}, where the focus is on detecting and characterizing rapid fluctuations in neural activity. The need for and impact of high-pass filtering are illustrated in Figure \ref{fig:normalized}. In this work, an FIR Least-Squares high pass filter is used, with a normalized cutoff of $0.043\pi$ rad/s.

\subsection{Voxel clustering}
\label{subsec: voxel clustering}

For each region of interest, voxel-level signals are clustered and selectively combined to obtain a single time-series representation of that region. Clustering \cite{clustering} identifies groups of voxels that share similar patterns, providing information on the temporal coincidence of regional activation. A phase-based clustering is performed which ensures that the activation within a region is not diluted or obscured by a smaller number of out-of-phase voxels. The phase of a voxel time-series signal is obtained by performing Discrete Fourier transform and ignoring the magnitude information. The K-means clustering algorithm is used on phase signals with $K = 3$. Post clustering, voxel signals within the cluster with the least intra-class variance are averaged to obtain a single time-series representation for each region. The least intra-class variance criteria ensures selection of signals which are coherent with each other and rejects out-of-phase signals. Functional connectivity among brain regions is then obtained by applying graph learning techniques on regional time-series data.

\begin{figure}
    \centering    
    \includegraphics[width=\linewidth]{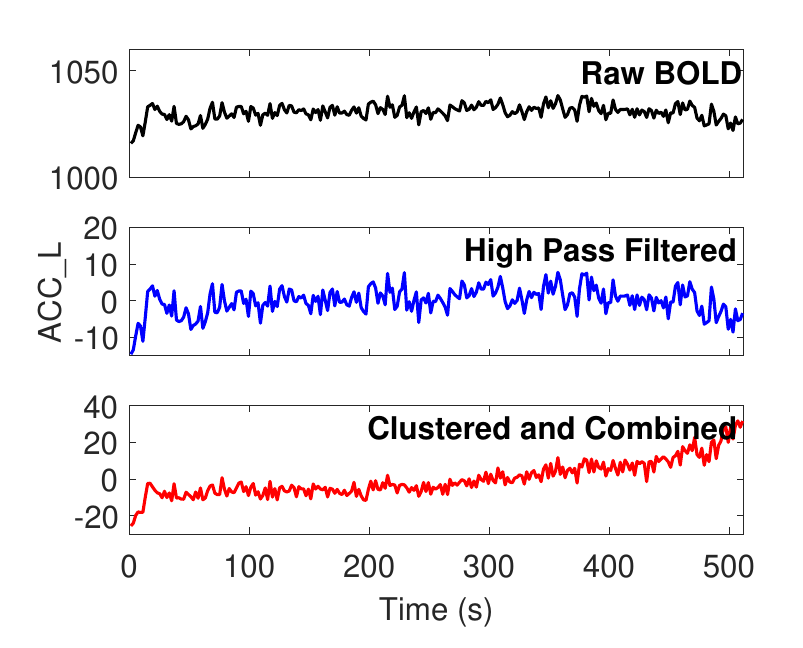}
    \caption{The progress in each step of the pipeline in processing the signal from raw BOLD to voxel-level combined BOLD.}
    \label{fig:pipelineProgress}
\end{figure}

So far, the pipeline focused on BOLD signal processing. Figure \ref{fig:pipelineProgress} illustrates the current status of the pipeline's progression. At each stage of the pipeline, all voxels time-series are combined to a single time-series through a simple average and subsequently analyzed. The impact of high-pass filtering is noticeable as it effectively eliminates slow variations in the signal attributed to factors such as scanner noise, breathing, and heartbeats—considered noise in the context of our task and thus rightfully discarded. The clustered and aggregated data visibly demonstrates that selecting voxels with similar phases enhances the capture of activations, preparing the data for the upcoming graph learning module. Subsequently, the upcoming module will involve the extraction of functional connectivity networks from the processed BOLD signals.

\subsection{Graph learning}
\label{subsec:GL}

In graph learning, graphs or networks $(\mathcal{G})$ are constructed from data $(\mathbf{X})$ which can help uncover hidden patterns and relationships among data points \cite{learningBasics}. Data points are represented as nodes in a graph, and edges are formed based on some notion of similarity measure between them. In our work, four graph learning techniques are used as discussed next.

\subsubsection{Pearson-coefficient}

{\bf Pearson-coefficient}-based graph learning \cite{pearson} method models the node signals as random vectors and finds the Pearson-correlation-coefficient between signals at every node pair. For $P$ dimensional signals, the Pearson-correlation-coefficient between the signals at nodes $n$ and $m$ is given as
\begin{equation}
    \rho(\mathbf{X}[n],\mathbf{X}[m]) 
    = \frac{\sum_{p=1}^{P}\left(\mathbf{X}_{p}[n]-\mu_{\mathbf{X}[n]}\right)\left(\mathbf{X}_{p}[m]-\mu_{\mathbf{X}[m]}\right)}{(P-1)\sigma_{\mathbf{X}[n]}\sigma_{\mathbf{X}[m]}}
\end{equation}
where, $\mathbf{X}_{p}[n]$ is the $p$th element of the signal at node $n$, $\mu_{\mathbf{X}[n]}$ and $\sigma_{\mathbf{X}[n]}$ are the mean and variance of the time-series signal $\mathbf{X}[n]$. This is directly used in the adjacency matrix construction by setting $\mathbf{A}_{nm} = |\rho(\mathbf{X}[n],\mathbf{X}[m])|$, since $\rho(.) \in [-1,1]$. Each graph signal is considered a random variable overlaid on the graph.

\subsubsection{Node-distance}

{\bf Node-distance}-based graph learning \cite{learningBasics} method computes the distance between the signals associated with every pair of nodes. The edge-weight $(\mathbf{A}_{nm})$ between the nodes $n$ and $m$ is then defined as
\begin{align}
\mathbf{A}_{nm} &= e^{-\frac{d_{nm}^{2}}{\sigma^{2}}} \,, \quad
d_{nm} = ||\mathbf{X}[n]-\mathbf{X}[m]||_{2} \,, \quad
\label{eq:similarity2}
\end{align}
where, $\mathbf{X}[n]$ and $\mathbf{X}[m]$ represent the graph signals at nodes $n$ and $m$ respectively, $||\cdot||_{2}$ represents the $\ell_{2}$ norm of a vector, $\sigma$ is a hyperparameter which controls the magnitude of weights generated. The choice of hyperparameters is provided in Section \ref{subsec:hyperparameter}. It is important to note here that the 'distance' referred to in the name of the method is not related to the anatomical distance between the regions, and is related to a distance-based metric between the signals ar the two regions.

\subsubsection{Sparsity}

{\bf Sparsity}-based graph learning \cite{learningBasics} tries to explain the signal at a given node using a sparse linear combination of signals at all the other nodes. Specifically, it solves an optimization problem for each node to extract its sparse connections which is converted into the adjacency matrix. For node $n$, the edge-weights $\boldsymbol{\beta}_{n} \in \mathbb{R}^{N-1}$ from node $n$ to all the other nodes $m = 1, 2, ..., n-1, n+1, ..., N$ are estimated by minimizing the following objective function
\begin{equation}
\label{eq:sparsity1}
    \hat{\boldsymbol{\beta}}_{n} 
    = \min_{\boldsymbol{\beta}_{n}}\ ||\mathbf{X}[n] - \mathbf{X}_{n}^{T}\boldsymbol{\beta}_{n}||_{2}^{2}
    + \lambda\ ||\boldsymbol{\beta}_{n}||_{1}
\end{equation}
where $\mathbf{X}_{n}$ is the data matrix of all the nodes except node $n$ and $||\cdot||_{1}$ represents the $\ell_{1}$ norm of a vector. The first term in the objective function ensures that the signal at node $n$ is well described by the chosen neighbors whereas the second term promotes sparsity of connections. $\lambda$ is a hyperparameter which controls the sparsity of the learnt edge-weights. The problem \eqref{eq:sparsity1} can be solved using the LASSO algorithm \cite{LASSO}. Since the solution generally does not provide a symmetric adjacency, it is obtained by performing
\begin{equation}
    \mathbf{A} = \sqrt{\mathbf{B}\mathbf{B}^{T}}
\end{equation}
where $\mathbf{B}$ is the non-symmetric adjacency constructed using $\boldsymbol{\beta}_{n}$ for all nodes $n$.

\subsubsection{Smoothness}

{\bf Smoothness}-based graph construction method learns a graph such that the given graph signals tend to be smooth on the learnt graph. It solves the following optimization problem
\begin{equation}
\label{eq:smoothness2}
    \{\hat{\mathbf{L}},\hat{\mathbf{Y}}\} = \min_{\mathbf{L}\in \mathcal{L},\mathbf{Y}}\ 
    \frac{1}{2} ||\mathbf{Y}-\mathbf{X}||_{F}^{2} + \alpha\ \text{Trace}\{{\mathbf{Y}}^{T}\mathbf{L}\mathbf{Y}\} + \beta\ ||\mathbf{L}||_{F}^{2}
\end{equation}
where, $||\cdot||_{F}$ denotes the Frobenius norm of a matrix and $\mathbf{L}$ is the graph Laplacian matrix. In \eqref{eq:smoothness2}, the first term represents data fidelity, which guarantees that the learned data ($\mathbf{Y}$) closely approximates the observed data ($\mathbf{X}$), the second term represents the smoothness of the data and the third term ensures the entries of the Laplacian are small. The optimization problem in \eqref{eq:smoothness2} is not convex in both $\mathbf{L}$ and $\mathbf{Y}$. This is addressed by solving for each of these variables separately in an iterative manner \cite{smoothness} as follows:
\begin{equation}
\label{eq:smoothness3}
    \hat{\mathbf{L}} = \min_{\mathbf{L}\in \mathcal{L}}\ \alpha\ \text{Trace}\{{\mathbf{Y}}^{T}\mathbf{L}\mathbf{Y}\} + \beta\ ||\mathbf{L}||_{F}^{2} \,,
\end{equation}
\begin{equation}
\label{eq:smoothness4}
    \hat{\mathbf{Y}} = \min_{\mathbf{Y}}\ \frac{1}{2} ||\mathbf{Y}-\mathbf{X}||_{F}^{2} + \alpha\ \text{Trace}\{{\mathbf{Y}}^{T}\mathbf{L}\mathbf{Y}\} \,.
\end{equation}

In \ref{eq:smoothness2} and \ref{eq:smoothness3}, $\mathbf{L}\in \mathcal{L}$ imposes a constraint that a valid Laplacian is learnt, where $\mathcal{L}$ denotes the set of all possible valid Laplacian matrices. In order to ensure this, the following two constraints are added to the optimization problem: $\mathbf{L}\mathbf{1} = \mathbf{0}$ and $\mathbf{L}_{ij} \leq 0\ \forall\ i\neq j,\ i,j=1,2,...,N$, which are the properties of a graph Laplacian matrix. Additionally, another constraint is added: $\text{Trace}\{\mathbf{L}\} = N$, which makes sure the trivial solution, that is, a null matrix is not obtained. The hyperparameter $\alpha$ governs the emphasis on smoothness, influencing the overall balance in the optimization process, while the hyperparameter $\beta$ dictates the magnitude of entries in the Laplacian matrix. Following graph construction, a thresholding process is applied, effectively utilizing $\beta$ to induce sparsity within the resultant graph. 

\subsection{Hyperparameter selection}
\label{subsec:hyperparameter}

The selection of hyperparameters in the graph learning methods is a critical step, and we conduct a thorough grid search to identify optimal values. In the node-distance-based graph learning method, the hyperparameter $\sigma$ is found to have an optimal value of 0.5, indicating its significance in shaping the learning process. For the sparsity-based graph learning method, the hyperparameter $\lambda$ is determined to be most effective at a value of 2.5, showcasing the importance of sparsity constraints in capturing meaningful relationships. The smoothness-based graph learning method involves a 2D grid search for hyperparameters $\alpha$ and $\beta$, with their optimal values identified as 0.25 and 9, respectively. This comprehensive exploration ensures that the chosen hyperparameter values contribute to the effectiveness of each graph learning method, providing insights into the underlying structure of the brain connectivity data.

\section{Analyses}
\label{sec:analyses}

The obtained graphs undergo further processing to extract essential metrics, considering two main dimensions: time and subjects. Temporal analysis is condensed into two distinct perspectives: 1) Averaging across all time points and 2) focusing on the time point with the highest emotional valence, labeled as "empathy HIGH". In terms of subjects, the analysis is performed by averaging across all subjects. The analyses are performed over both the movies separately. The analyses include graph cluster labels, connectome-network analysis, edge-weight dynamics, and region identification using spectral filtering, detailed in the subsequent subsections. In all the analyses, due to the absence of any concrete ground truth of the functional connectivity networks, performance analysis has been done by comparing the emotion scale with secondary metrics extracted from the obtained networks, like graph cluster labels, edge-weight dynamics and region graph signal values in spectral filtering.



\subsection{Graph clustering-based analysis}

After generating graphs with the graph learning module, the graphs are clustered as detailed in Section \ref{subsec:graph clustering}.Despite the current utilization of a binary clustering algorithm, which simplifies the identification of empathy presence at a given time instant, this choice proves effective in discerning optimal graph learning methods for the specific stimulus, which is the primary objective of this analysis. The evaluation focuses on their ability to distinguish between empathy-related and non-empathy-related networks, aligning with the emotion scale. The diverse assumptions inherent to different graph learning methods come to light through this analysis, aiding in the formulation of informed decisions regarding the properties and assumptions applicable to brain signals, particularly concerning connectivity during an external stimulus. Subsequently, the empathy scores are generated and averaged across subjects as detailed in Section \ref{subsec:cross-corr & score}, which serve as a foundational step in identifying the most effective graph learning method.

\subsection{Edge-weight dynamics}

This analysis comprises two distinct sub-parts: 1) Temporal evaluation of edge-weight between specific regions associated with empathy, and 2) analysis of edge-weight dynamics in relation to the emotion scale. After obtaining the graphs, the focus shifts to extracting the edge-weight between two regions and analysing its temporal dynamics. In the first segment, we investigate the simultaneous activation of two empathy-specific regions, as reflected in their edge-weights over time. From this analysis, we aim to highlight the impact of the sparsity-based approach over the others in capturing temporal coincidences in region activations.

The second segment involves analyzing the dynamics of edge-weights for all possible connections within the obtained graphs. Our objective is to identify the top 5 edges that exhibit the highest alignment with the emotion scale, facilitating the recognition of edges crucial to empathy. The alignment is quantified using the cross-correlation metric. This approach aids in comprehensively understanding the interplay between specific regions and identifying edges that strongly correlate with the emotion scale, thereby highlighting connections integral to an empathic response.

\subsection{Connectome-based network analysis}

As a part of this analysis, edge thresholding is implemented on the obtained graphs, specifically focusing on the top 5 edges while discarding the remaining ones. This process allows us to highlight networks within the brain, emphasizing significant and strong connections during distinct time intervals throughout the stimulus. The primary objective is to delineate and compare the isolated regions resulting from diverse time-based analyses. This comparison becomes particularly vital as it enables us to contrast connectivity patterns and identify regions characterized by robust connections, differentiating between all-time-average and the empathy HIGH time-window.

\subsection{Region identification with spectral filtering}

This analysis delves into the GFT-based filtering method outlined in Section \ref{subsec: GFT}. Notably, the smoothness-based graph learning approach carries an inherent assumption of data smoothness across the learned graph. Consequently, studying the spectral decomposition of data over such graphs may be less suitable, given that energy tends to concentrate primarily on the lower set of graph frequencies. In contrast, sparsity-based graph learning avoids such assumptions, making its spectral analysis an insightful exploration across various frequency bands. This study specifically focuses on three bands: 1) the low-pass band (comprising the bottom one-third of the graph frequency spectrum), 2) the band-pass band (encompassing the middle one-third) and 3) the high-pass band (encompassing the upper one-third). As detailed in Section \ref{subsec: GFT}, emphasis is placed on the band-pass band due to its capacity to concentrate on specific patches and sets of regions within the graph, in contrast to the broad coverage of the low-pass band or the deeper localization inherent in the high-pass band.

The procedure unfolds step-wise, commencing with the graph Fourier transform applied over the graph signal at a chosen time window with respect to its corresponding graph. Given that the graphs are learned in a windowed manner, the data is also segmented into windows, with the median of each window selected for the graph spectral analysis. The resulting graph Fourier transform coefficients act as carriers of energies of the specific signal at the corresponding graph frequencies, akin to classical Fourier transform principles. Subsequently, band-pass filtering is executed by discarding the lowest one-third and highest one-third coefficients, retaining only the band-pass band. These coefficients are then utilized to perform an inverse graph Fourier transform, ultimately transforming back to the node-domain signal.

The primary objective of this analysis is to scrutinize groups of regions associated with empathy-specific activations and analyse variations highlighted across different time-based analyses. Notably, this analysis extends beyond the realm of BOLD signal activations, capturing nuances attributable to the underlying graph structure. This multifaceted approach adds depth to the exploration of empathy-specific regions, contributing valuable insights to our understanding of the neural correlates of empathy.

\section{Results}
\label{sec:results}

\begin{figure}
    \centering    
    \includegraphics[width=0.9\linewidth]{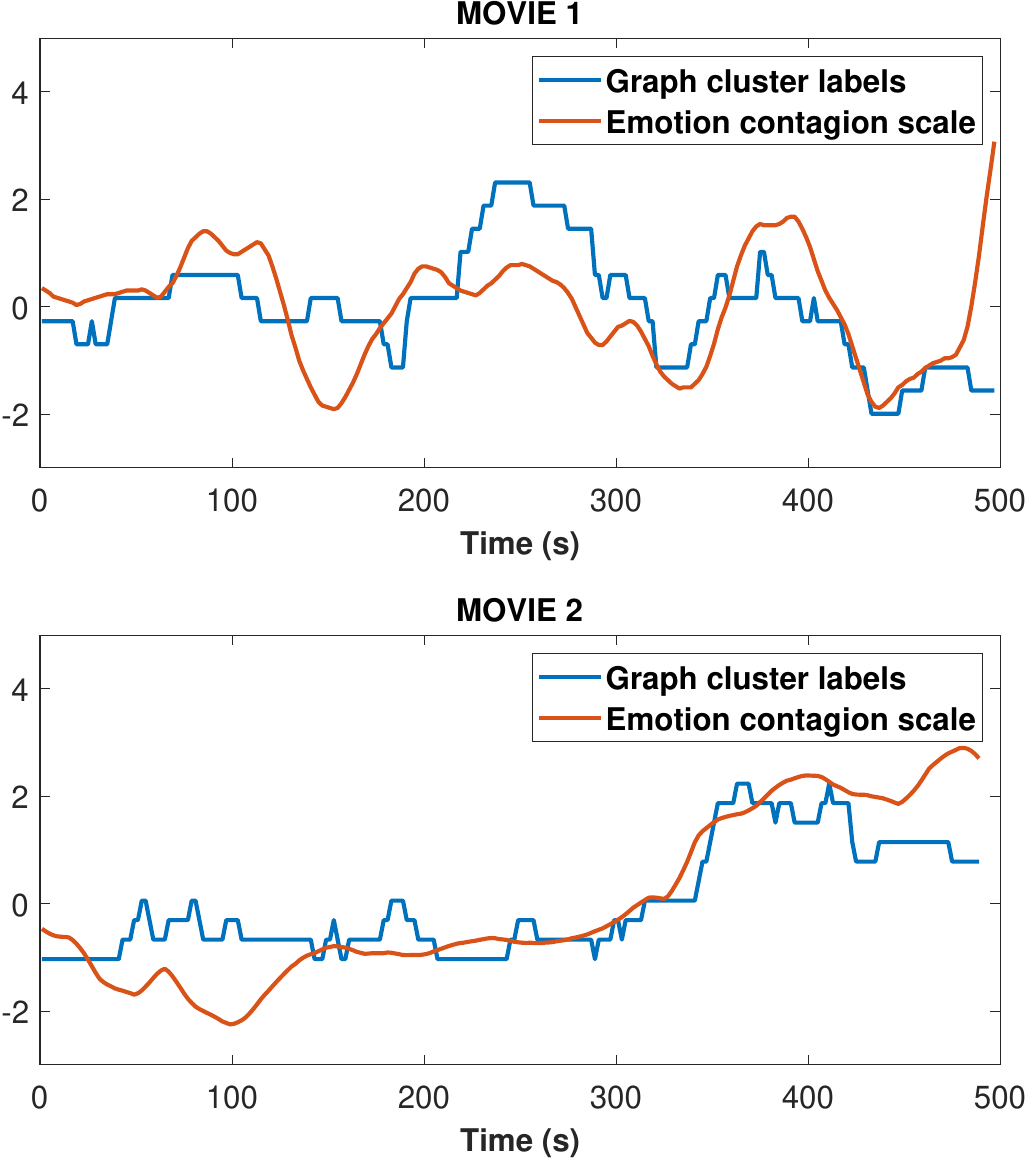}
    \caption{The subject-averaged graph cluster labels superimposed with emotion contagion scale plotted over time for movie 1 (top) and movie 2 (bottom) for the sparsity-based approach}
    \label{fig:labelVsScale}
\end{figure}

To establish a foundational framework for our results, we initiate with a comprehensive analysis across different graph learning methods, with the objective of identifying the most effective graph learning method based on a specific metric. Our choice of metric involves utilizing average empathy scores, which are the correlation percentages between graph clustering labels and the emotion scale, averaged across subjects. The subsequent subsection, Section \ref{subsec:graph clustering result}, reveals the exceptional performance of the sparsity-based method in aligning with the emotion scale, in both the movies. Consequently, we restrict our subsequent analyses exclusively to sparsity-based graph learning. Nevertheless, to provide a comparative perspective, we incorporate a thorough and extensive study comparing results across various methods in Annexure C. Moreover, results with all-time-based analyses like graph cluster labels and edge-weight dynamics are excluded from the results with the two broad time-based analyses as they require the whole time-series. On the other hand, the remaining analyses, such as connectome analysis and region identification with spectral filtering, undergo both the time-based analyses, and the outcomes for movie 1 are systematically presented in Table \ref{tab:region summary}. All the analyses are performed for both the movies, the edge-weight dynamics result for movie 2 is not presented here as it shows similar results across both movies.



\subsection{Graph clustering vs emotion contagion scale}
\label{subsec:graph clustering result}

\begin{figure}
    \centering \includegraphics[width=\linewidth]{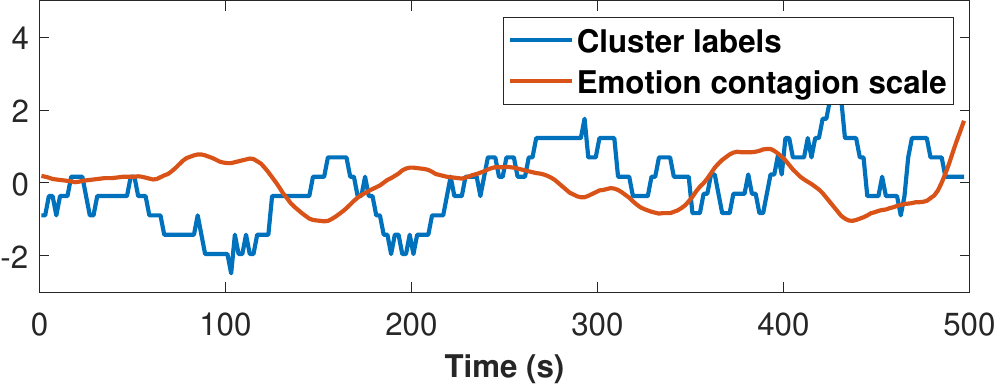}
    \caption{The subject-averaged graph cluster labels superimposed with emotion contagion scale plotted over time for Pearson's correlation-based approach for movie 1}
    \label{fig:labelVsScalePearson}
\end{figure}

The sparsity-based approach consistently captures variations in the emotion contagion scale as shown in Figure \ref{fig:labelVsScale}, with all participants demonstrating alignment of over 80\% with the emotion scale. When subject responses are averaged, the distance, Pearson, smoothness, and sparsity methods exhibit 80\%, 72\%, 80\%, and 88\% match with the emotion scale, respectively. Figure \ref{fig:labelVsScale} (top) illustrates the temporal variation of graph cluster labels averaged across all subjects overlaid on the emotion scale for movie 1.

A notable finding is that, up to the 180s mark in the movie, the averaged graph clustering labels show limited correlation with the emotion scale. However, they become highly synchronized post 180s. The above mentioned values are with respect to movie 1. Figure \ref{fig:labelVsScale} (bottom) shows temporal variation of graph cluster labels averaged across all subjects overlaid on the emotion scale for movie 2, and it can be seen that similar to movie 1, sparsity-based approach is able to consistently capture emotion scale variations after 120s, with an overall of 88\% match. 

As a comparison with the existing methods, Figure \ref{fig:labelVsScalePearson} shows that the Pearson's correlation-based approach fails to capture temporal variations, with a 72\% match with the emotion scale, showing the ability of sparsity-based approach to capture activations better in a task-specific stimulus.

\subsection{Edge-weight dynamics}

\begin{figure}
    \centering    
    \includegraphics[width=\linewidth]{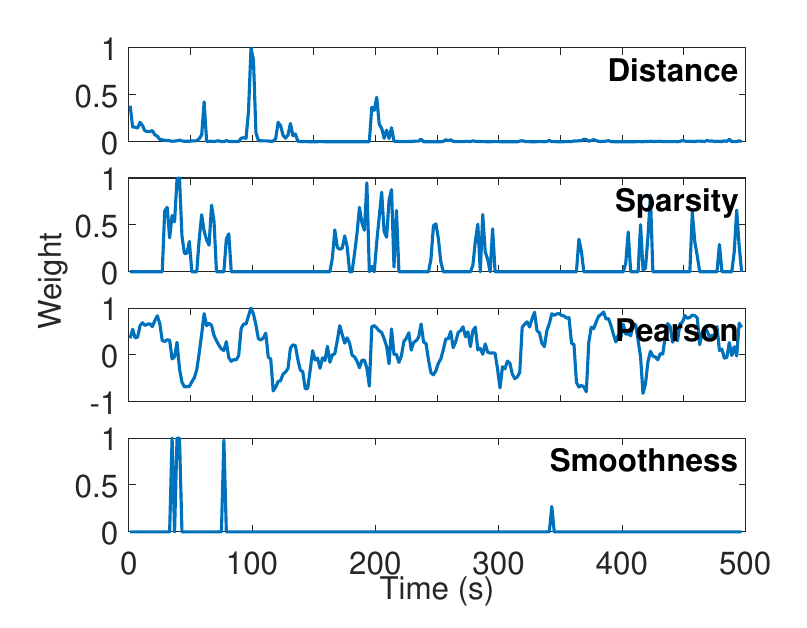}
    \caption{Edge-weight between Insula\_R and Opercularis\_R for one subject using all methods}
    \label{fig:303RR}
\end{figure}

The time dynamics of the edge-weight between Insula\_R - Triangularis\_R was analysed. Figure \ref{fig:303RR} compares different graph learning methods for a randomly selected subject, where the distance method exhibits peaks around 250s in the Insula-Triangularis connection for 10 out of 14 subjects, while the Pearson method lacks clear edge identification. The sparsity-based method identifies peaks around 250s and 475s, which aligns with the peaks observed in the emotion scale in Figure \ref{fig:labelVsScale}. Figure \ref{fig:303RR} distinctly illustrates the susceptibility of Pearson's correlation-based approach to noise, resulting in noisy activations. In contrast, both the distance and smoothness methods exhibit limited or negligible activations. This underscores the superior performance of the sparsity-based approach when dealing with a noisy, task-based stimulus, with activations in-line with the emotion scale.

In the analysis targeting edges with edge-weight dynamics most correlated with the emotion scale, several significant connections were identified. Averaging across all subjects revealed prominent edges, including Frontal\_Sup\_Medial\_L $\leftrightarrow$ Frontal\_Sup\_R, Frontal\_Inf\_Orb\_R $\leftrightarrow$ Temporal\_Inf\_L, ParaHippocampal\_L $\leftrightarrow$ Temporal\_Inf\_L, Frontal\_Inf\_Orb\_R $\leftrightarrow$ Temporal\_Inf\_R, and Frontal\_Med\_Orb\_R $\leftrightarrow$ Temporal\_Inf\_R for movie 1. The objective of conducting this analysis was to illustrate the impact of the sparsity-based approach compared to other methods. Consequently, the analysis was exclusively carried out on movie 1 and not presented for movie 2. 

\subsection{Connectome analysis}
\label{subsec:connectome result}

\begin{figure*}
    \centering
     \begin{subfigure}{0.48\textwidth}
         \centering
         \includegraphics[width=\linewidth]{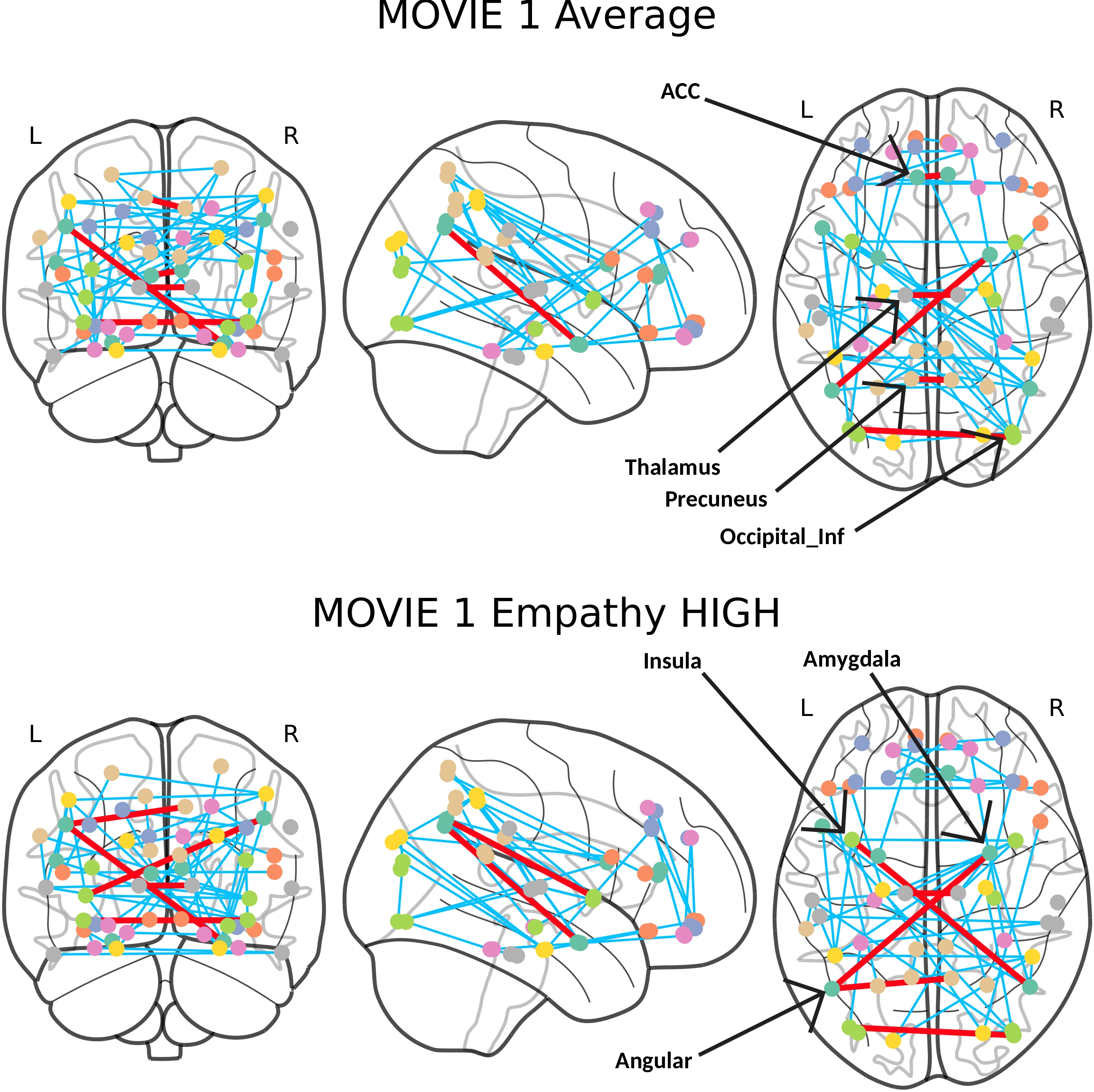}
         \caption{ }
         \label{fig:connectome1}
     \end{subfigure}
    \hfill
    \begin{subfigure}{0.48\textwidth}
        \centering
        \includegraphics[width=\linewidth]{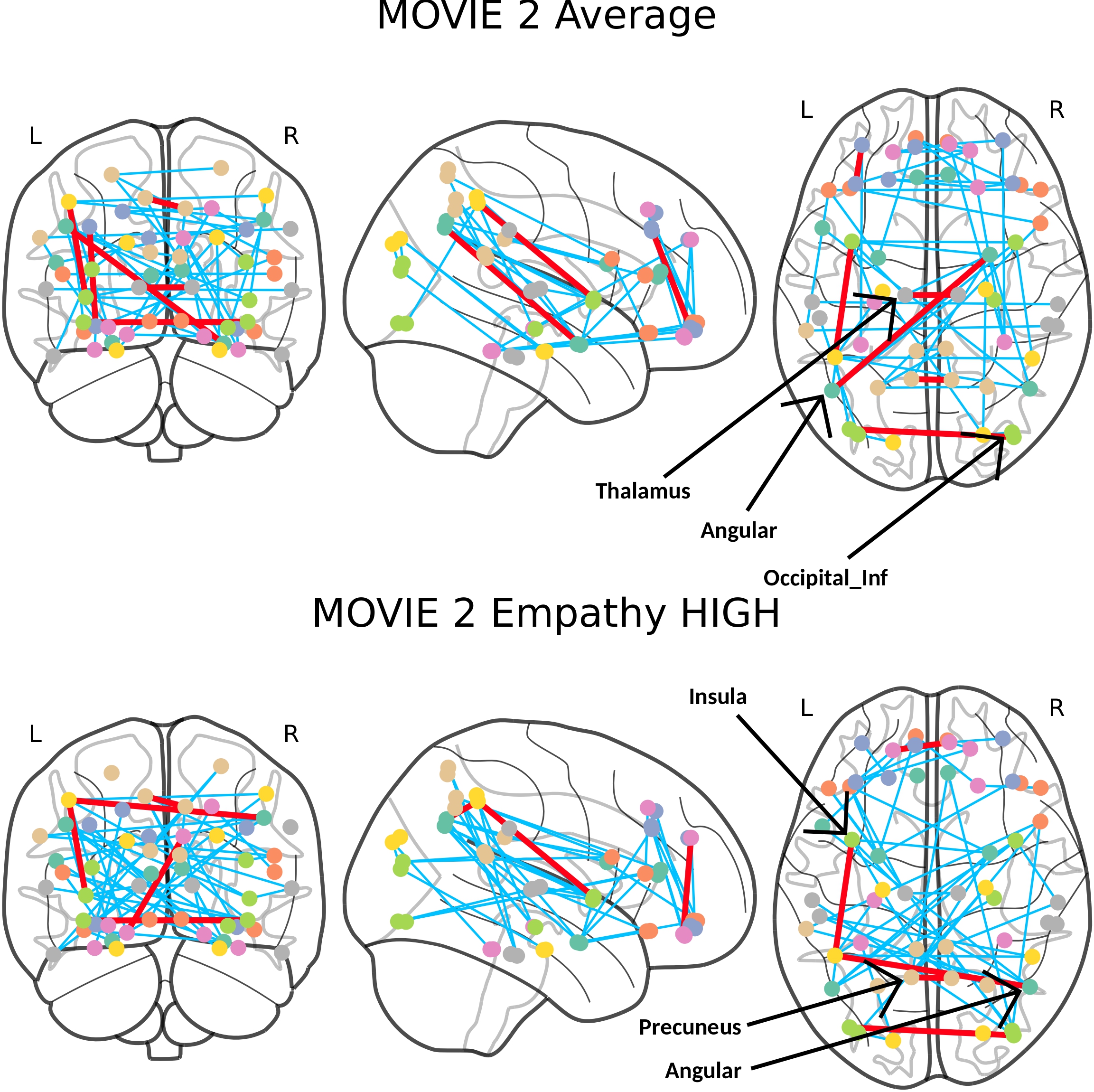}
        \caption{ }
        \label{fig:connectome2}
    \end{subfigure}
    \caption{Connectome graph illustrating the networks learnt using sparsity-based graph learning. The significant nodes are labelled according to the AAL Atlas. The blue edges correspond to the significant top-$N$ edges, and the red edges correspond to the strongest 5 edges in the graph Adjacency. It shows variation in networks obtained by averaging on all time (top) and during empathy HIGH (bottom) in movie 1 (left) and movie 2 (right)}
\end{figure*}

The listed networks comprise a collection of regions characterized by the strongest edges within the acquired graphs, as shown in Figure \ref{fig:connectome1}. This selection is based on the top 5 edges (in magnitude). The two regions associated with each of these edges are listed. In the context of connectome-based network analysis, specifically during the empathy HIGH time-window, the regions Insula\_L, Parietal\_Inf\_R, Thalamus\_L and Amygdala\_R consistently emerge. In the scenario of all-time-average analysis, we observe bi-lateral connections being formed with the same region for both the movies, specifically between ACC\_L - ACC\_R, Thalamus\_L - Thalamus\_R, and Precuneus\_L - Precuneus\_R. During all-time-averaged analyses across both the movies, we observed that the edges between Amygdala\_R - Angular\_R and Occipital\_Inf\_L - Occipital\_Inf\_R prevail. During the empathy HIGH time-window, in both the movies, the edge between Occipital\_Inf\_L and Occipital\_Inf\_R is not present. Summarizing, across both the movies, during the empathy HIGH time window, the Insula, Amygdala and ACC and Angular appear to have strong connections.

\subsection{Region identification with spectral filtering}

\begin{figure}
    \centering \includegraphics[width=\linewidth]{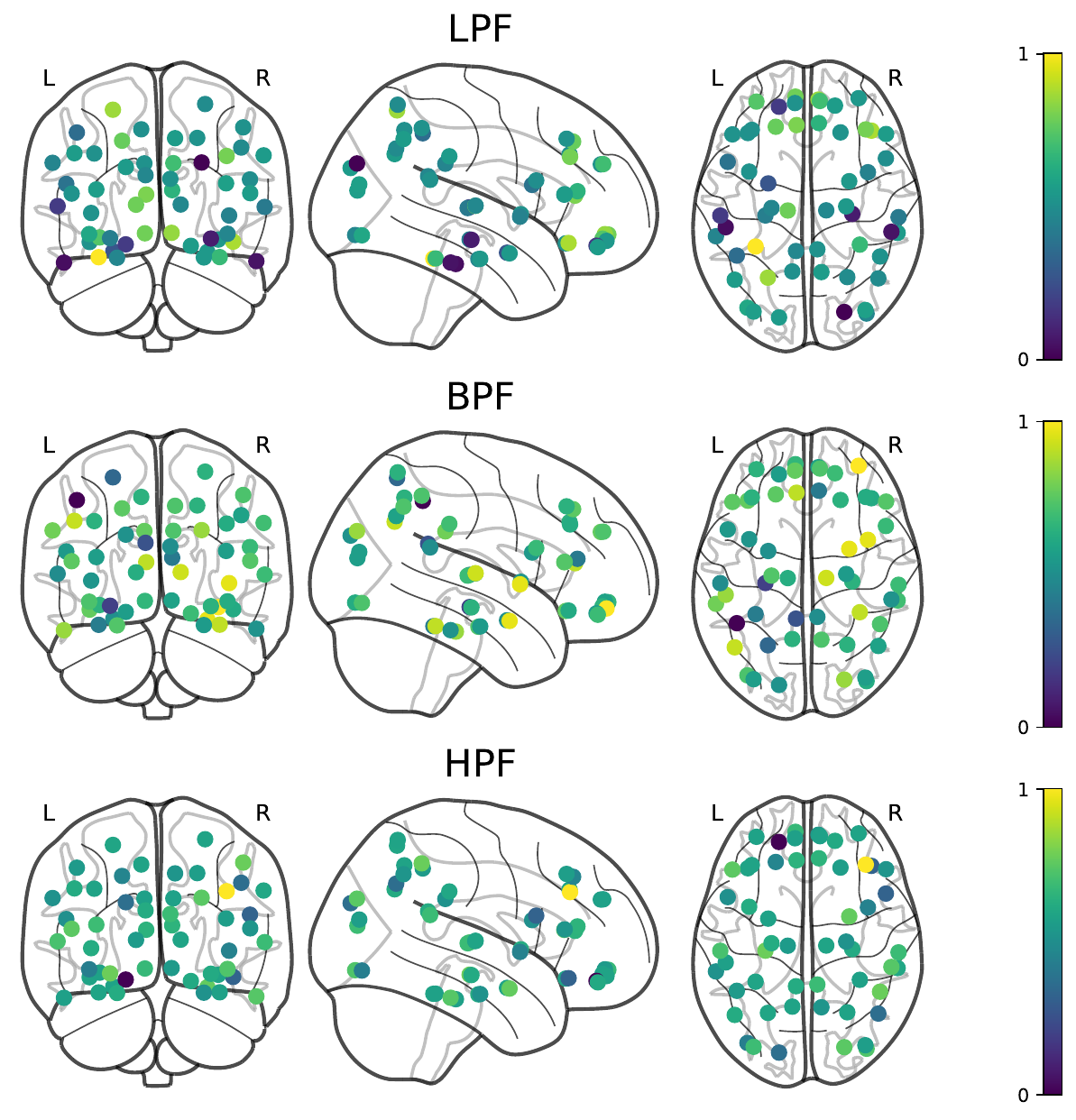}
    \caption{GFT of the graph signal at empathy HIGH time window, filtered into all 3 bands using sparsity-based approach averaged across all subjects.}
    \label{fig:AllBands}
\end{figure}

\begin{figure*}
    \centering
     \begin{subfigure}[t]{0.48\textwidth}
         \centering
         \includegraphics[width=\linewidth]{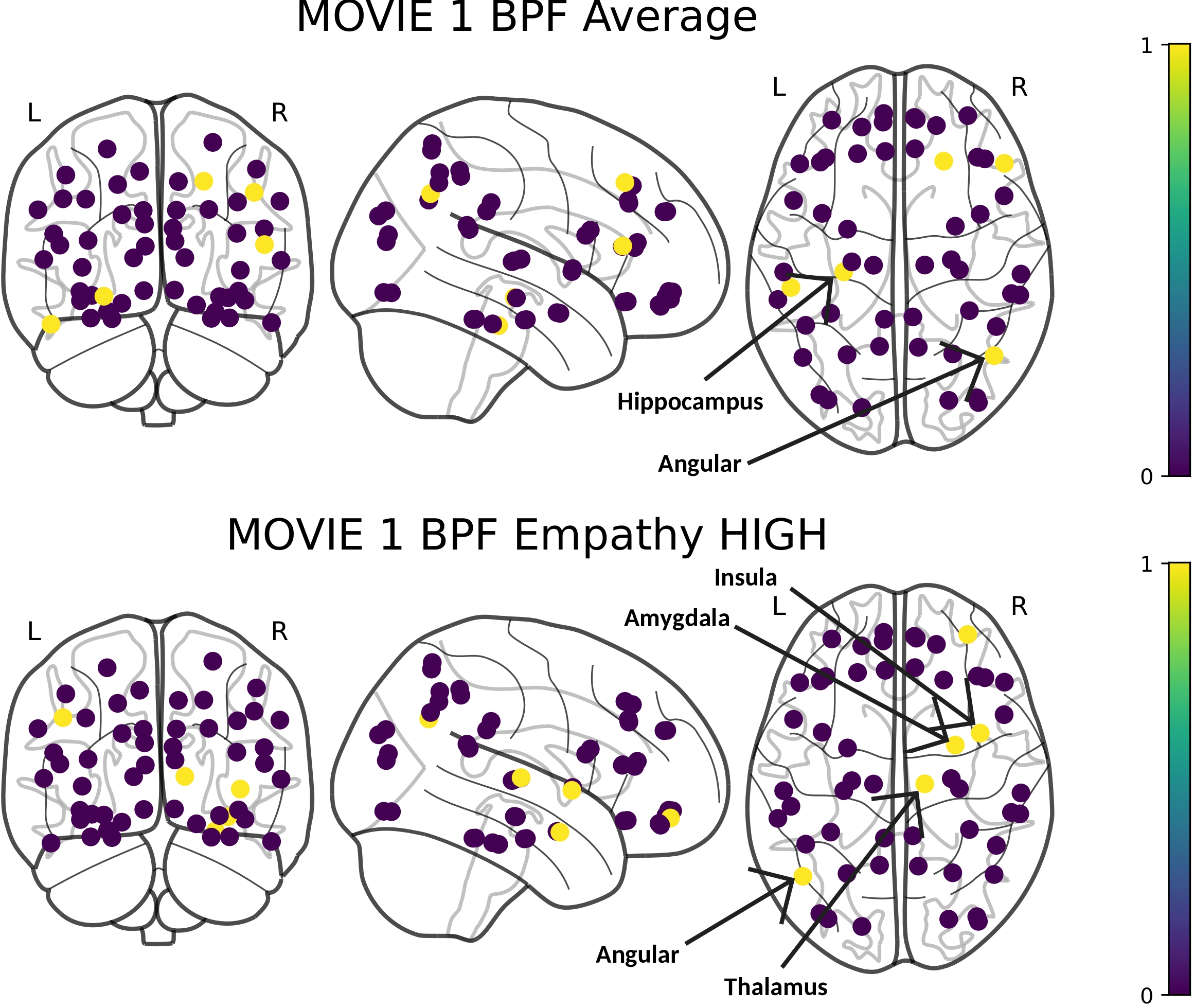}
         \caption{ }
         \label{fig:gft1}
     \end{subfigure}
    \hfill
    \begin{subfigure}[t]{0.48\textwidth}
        \centering
        \includegraphics[width=\linewidth]{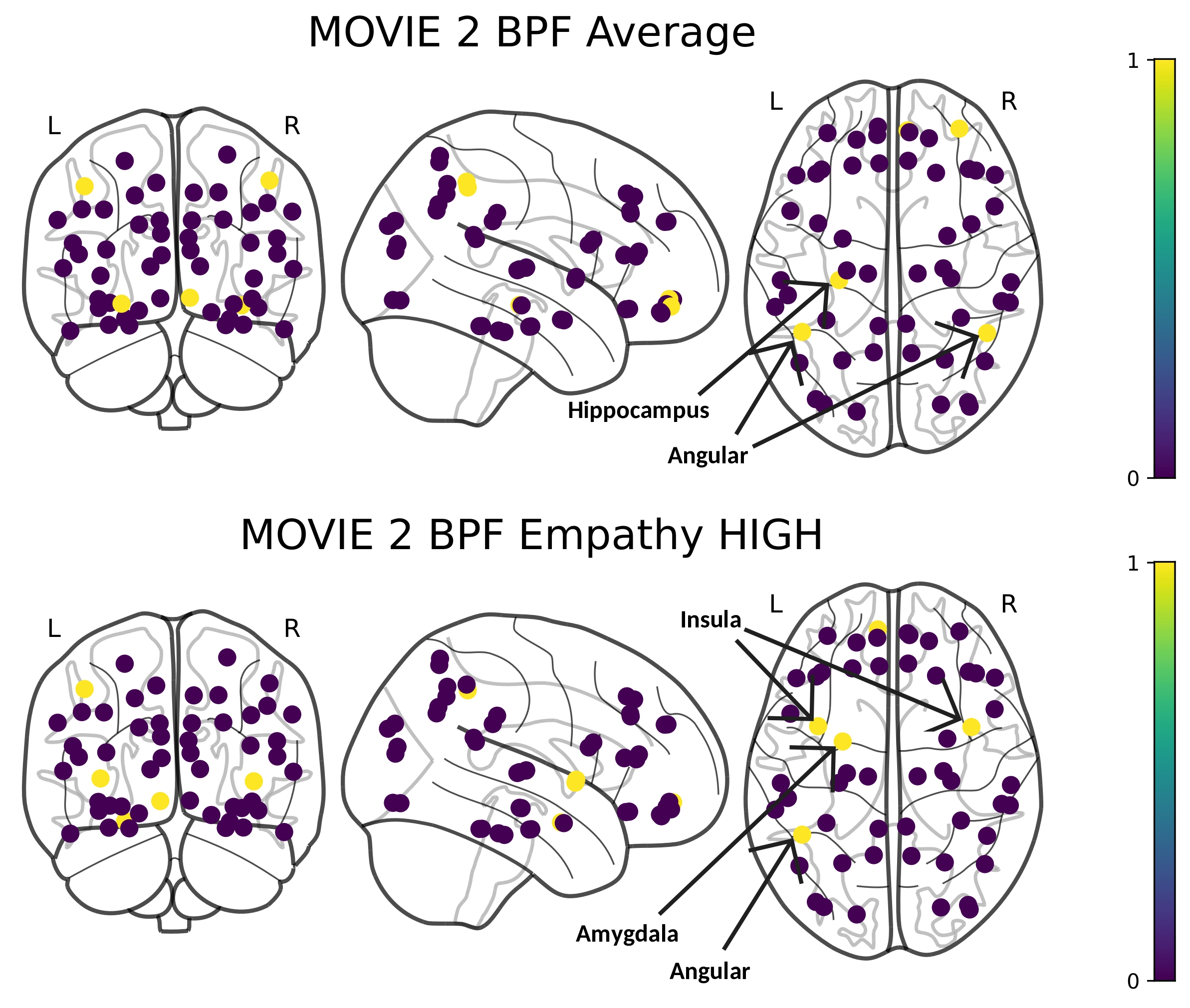}
        \caption{ }
        \label{fig:gft2}
    \end{subfigure}
    \caption{Region plot with the regions identified with with spectral filtering (BPF) using sparsity-based approach. The top 5 regions are highlighted. It shows variation in regions highlighted by averaging on all time (top) and during empathy HIGH (bottom) in movie 1 (left) and movie 2 (right)}
\end{figure*}

\begin{table}[t]
\centering
\begin{tabular}{|p{1.5cm}|p{2cm}|p{4cm}|} 
\hline
Analysis & Time & Avg. on all subjects \\
\hline
\multirow{3}{4em}{Connectome - network} & Avg. on all time & ACC\_L $\leftrightarrow$ ACC\_R, Thalamus\_L $\leftrightarrow$ Thalamus\_R, Precuneus\_L $\leftrightarrow$ Precuneus\_R, Amygdala\_R $\leftrightarrow$ Angular\_L\\ \cline{2-3}
 & Empathy HIGH & Insula\_L $\leftrightarrow$ Parietal\_Inf\_R, Thalamus\_L $\leftrightarrow$ Thalamus\_R, Occipital\_Inf\_L $\leftrightarrow$ Occipital\_Inf\_R, Amygdala\_R $\leftrightarrow$ Angular\_L\\ \cline{2-3}
\hline
\multirow{3}{4em}{Spectral filtering} & Avg. on all time & Hippocampus\_L, Angular\_R, Frontal\_Inf\_Tri\_R, Frontal\_Sup\_R, Temporal\_Inf\_L\\ \cline{2-3}
 & Empathy HIGH & Amygdala\_R, Insula\_R, Thalamus\_R, Angular\_L, Frontal\_Mid\_Orb\_R\\ \cline{2-3}
 \hline
\end{tabular}
\caption{Summary of regions and networks identified through graph Fourier transform-based region identification and connectome-based network analysis for different empathy levels for movie 1}
\label{tab:region summary}
\end{table}

In contrast to the emphasis on edges in connectome-based network analysis, the spectral filtering-based approach prioritizes the identification of regions across different graph frequency bands. Signals that have undergone spectral filtering are visualized over the graph, and the regions exhibiting the top 5 amplitudes are selected.

Specifically, during GFT-based filtering during the empathy HIGH time window, pivotal regions such as Amygdala\_R, Insula\_R, Thalamus\_R, Angular\_L, and Frontal\_Mid\_Orb\_R consistently stand out for movie 1. These regions are recurrently highlighted across the top 5 amplitudes, underscoring their significance in capturing heightened empathy states. In contrast, during all-time-average analysis, distinctive regions, including Frontal\_Sup\_R, Hippocampus\_L, Temporal\_Inf\_L, Angular\_R, and Frontal\_Inf\_Tri\_R (Triangularis), emerge prominently, demonstrating shifts in neural activity associated with empathy states on an average across time.

Regarding movie 2, in the empathy HIGH time window, common highlighted regions include Amygdala\_L, Insula\_L, and Insula\_R, mirroring the observations from movie 1. Additionally, Fusiform\_L and Frontal\_Med\_Orb\_L emerge as noteworthy regions uniquely emphasized during this period. In Figure \ref{fig:AllBands}, the graph signal at each region is depicted on a colorbar. The loww-pass (LPF) band prominently showcases smooth variations across the graph, with connected regions exhibiting similar node colors. The high-pass (HPF) band brings attention to deeper variations, emphasizing specific regions with distinct values from their connected counterparts. Meanwhile, the babd-pass (BPF) band effectively highlights patches, aligning with the anticipated outcome.

\section{Discussion}
\label{sec:discussion}

Graph theory efficiently assesses brain network states, modeling inter-relationships between brain areas via edges and nodes using various metrics. Previous studies have applied graph theory to identify measures of neurological and psychiatric disorders \cite{Basset2,Zhang1}, emotional brain states, and task-aware effective brain connectivity \cite{GNN1}. However, it hasn't been utilized to study empathy-specific regions to the best of our knowledge. Hence, we employed graph learning methods for whole-brain analysis to investigate brain activity in response to a naturalistic stimuli, expecting empathic responses from participants.

Using GSP for fMRI empathy analysis offers several advantages. GSP captures complex brain network interactions, providing a deep understanding of functional connectivity and dynamics during empathetic processes. It uncovers hidden patterns not easily discernible with traditional methods, potentially leading to novel insights. GSP's adaptability to individual differences accommodates inter-subject variability, allowing for both group and individual-level empathy-related brain activity analysis. Please note that in all the analyses, due to the absence of any  ground truth for the functional connectivity networks, performance analysis has been done by comparing the emotion scale (behavioural) with secondary metrics extracted from the obtained networks, like graph cluster labels, edge-weight dynamics and region graph signal values in spectral filtering.

The sparsity-correlation-based graph learning approach is well-suited for empathy-related tasks due to its capacity to efficiently capture selective and specific interactions among brain regions. This method adapts to individual differences, effectively representing dynamic and localized neural activations associated with empathy. Its emphasis on temporal dynamics aligns with the rapid fluctuations inherent in empathy-related processes. The resulting sparse graphs enhance interpretability, reducing the susceptibility to noise, offering a focused and biologically plausible representation of the neural network. By suppressing non-specific connections, this approach enhances sensitivity to task-relevant signals, providing a powerful tool for investigating the nuanced and transient nature of the neural correlates of empathy.

The sparsity-based approach consistently captures variations in the emotion contagion scale (Fig. \ref{fig:labelVsScale}), demonstrating an alignment of over 80\% with the emotion contagion for all participants. When averaging subject responses, the distance, Pearson, smoothness, and sparsity-based methods exhibit 80\%, 72\%, 80\%, and 88\% match with the emotion scale, respectively. In comparison, Figure \ref{fig:labelVsScalePearson} demonstrates that the Pearson's correlation-based approach fails to capture temporal variations, achieving a 72\% match with the emotion scale. This highlights the superior ability of the sparsity-based approach to capture activations in a task-specific stimulus over existing methods, in both the movies.

In the observed graph clustering dynamics with the sparsity-based approach, a notable misalignment among clusters is discernible before the 180-second mark, suggesting a lack of synchronization. However, post the initial period, a strong alignment is evident in both the movies, indicating a coherent configuration of clusters. The observed temporal shift in cluster alignment can be attributed to the gradual induction of empathy into the subjects rather than occurring suddenly, indicating a slow and steady build-up of empathetic responses. Notably, as empathy peaks in the subject, a synchronous alignment between the graph clusters and the emotion scale becomes evident in both the movies. This alignment provides compelling evidence that the obtained graph structures indeed correspond to an empathy network, further substantiating the temporal relationship between empathy induction and the configuration of functional brain networks.

Analyzing edge-weight variations between the Insula and Triangularis over time reveals that sparsity-based learning consistently detects activations more frequently than smoothness-based learning, particularly around the 200s mark. This indicates that smoothness-based learning may lead to fewer activations, especially in the Insula, due to its uniform activation assumption. Sparsity-based learning excels in identifying influential neighboring nodes, offering a more insightful perspective. Both methods exhibit peaks around 475s, requiring further investigation for nuanced interpretation, potentially linked to emotionally weighted movie scenes. 

Analysing distance-based and smoothness-based methods reveals limited activations, reducing sensitivity, while Pearson's correlation-based method produces noisy activations, hindering clear pattern understanding. This highlights the superior performance of the sparsity-based method in capturing relevant activations in empathy-related neural processes. In comparison with other methods concerning edge-weight activations and graph cluster alignment, the sparsity-based approach appears to outperform. Table \ref{tab:method comparison} provides a tabulated summary of these results with existing methods.

\begin{table}[t]
\centering
\begin{tabular}{|p{1.8cm}|p{1.2cm}|p{1.4cm}|p{2.5cm}|} 
\hline
Method & Avg. empathy score & Edge-weight activations & Major regions \\
\hline
Distance \cite{Huang1} & 82\% & Too-few & \textbf{Supramarginal\_L}, Parietal\_Sup\_R, Parietal\_Sup\_R, Occipital\_Inf\_L\\
\hline
Pearson's Correlation \cite{pearson} & 72\% & Too-many, noisy & Frontal\_Med\_Orb\_L, \textbf{Fusiform\_R}, Occipital\_Inf\_L, Occipital\_Inf\_R\\
\hline
Smoothness \cite{smoothness} & 82\% & Too-few & Frontal\_Mid\_Orb\_R, Angular\_R, Parietal\_Sup\_R, Temporal\_Inf\_R \\
\hline
\textbf{Sparsity} & \textbf{88\%} & \textbf{Ideal} & \textbf{Amygdala\_R}, \textbf{Insula\_R}, \textbf{Thalamus\_R}, Angular\_L, Frontal\_Mid\_Orb\_R \\
\hline
\end{tabular}
\caption{Comparison of proposed pipeline of sparsity-based method with existing literature.}
\label{tab:method comparison}
\end{table}

In analysing edge-weight dynamics aligned with the emotion scale, key edges consistently activated during empathetic responses emerge as crucial markers, particularly in frontal and temporal regions, with notable emphasis on the Parahippocampal area for movie 1. Given the Parahippocampal region's well-established role in episodic memory and retrieval, its heightened activation aligns with the potential triggering of memory-based events during emotionally weighted movie scenes. This underscores the significance of the Parahippocampal edge in the intricate interplay between empathetic responses and episodic memory retrieval.

The connectome-network analysis identifies the Insula, Amygdala, and Thalamus as central regions in empathy, aligning with established neuroscientific roles \cite{Lamm1}. The Insula integrates emotional experiences, the Amygdala regulates emotions, and the Thalamus processes emotional cues. Lateral brain connections within homologous regions suggest synchronized responses, facilitated by the corpus callosum. This inter-hemispheric coordination enhances the brain's holistic processing of empathetic information. Averaging across all time reveals heightened activity in the ACC, Thalamus, and Precuneus. ACC's role in the default mode network, Thalamus' sensory relay, and Precuneus' affective responses contribute to continuous engagement. During empathy HIGH, Insula, Angular Gyrus, and frontal regions show prominent connections. As a part of our study, Angular Gyrus appears to get activated in correlation with the known empathy areas like the Amygdala in both the movies, highlighting its pivotal role in heightened empathy periods.

As a part of the spectral filtering-based analysis, Figure \ref{fig:AllBands} displays patterns in low-pass (LPF), high-pass (HPF) and band-pass (BPF) bands, offering unique insights. LPF emphasizes stability in specific nodes, while HPF highlights anomalies, proving ineffective in isolating empathy-specific networks as it consists of multiple regions with correlated activations. Notably, BPF strikes a balance between stability (almost constant throughout the regions) and anomaly (deeper localization), thereby isolating regions linked to emotional and empathetic processing during the empathy HIGH state. Key regions, including Amygdala\_R, Insula\_R, Thalamus\_R, Angular\_L, Frontal\_Mid\_Orb\_R, exhibit heightened activations in both the movies, emphasizing their role in immediate emotional responses. In the all-time-average analysis, sustained activations in Hippocampus\_L, Angular\_R, Frontal\_Inf\_Tri\_R (Triangularis), Frontal\_Sup\_R, Temporal\_Inf\_L suggest broader contributions to empathy across the entire stimulus duration. The activation in the Hippocampal area suggests memory retrieval taking place throughout the movie along with highlighting its contributions in emotion processing \cite{Amygdala2}. These findings align with literature \cite{Lamm1,Amygdala2,Amygdala1,Insula}, on neural correlates of empathy, reinforcing the selected BPF band as optimal for capturing empathy-related neural activity.

Both movies exhibited strong similarities across various analyses, ranging from similar patterns in graph cluster labels matching the emotion scale, to consistent highlighting of regions like Insula, Amygdala, and Angular in connectome-network and spectral filtering-based analyses. While the obtained regions are largely similar, potential differences in cognitive processing associated with each movie could contribute to subtle distinctions. However, regions highlighted in common can be claimed to a stronger foundation in eliciting empathetic responses in the brain.

\section{Conclusion}
\label{sec:conclusion}

In this paper, we present a processing pipeline to extract dynamic functional connectivity patterns between brain areas through the sparsity-based graph learning method. The sparsity-based approach consistently outperforms other methods in capturing variations in the emotion contagion scale, aligning with over 80\% accuracy across all participants. The temporal shift in graph cluster alignment indicates a gradual induction of empathy, supporting the method's effectiveness in capturing dynamic connectomes. Edge-weight dynamics analysis reveals the superiority of sparsity-based learning, particularly in detecting activations around the 200s mark, emphasizing its ability to identify influential neighboring nodes. 

A comprehensive method comparison highlights sparsity's ideal performance in empathy HIGH state, outperforming existing methods in terms of average empathy scores, edge-weight activations, and major regions involved. Connectome-network analysis underscores the pivotal role of the Insula, Amygdala, and Thalamus in empathy, with lateral brain connections facilitating synchronized responses. Spectral filtering analysis demonstrates the significance of the band-pass filter in isolating regions linked to emotional and empathetic processing during empathy HIGH states. The consistent activation of key regions like Amygdala, Insula, and Angular Gyrus further supports their critical role in immediate emotional responses. Overall, the obtained results across two movies reveal strong similarities in graph cluster labels, connectome-network analysis, and spectral filtering-based analyses, indicating robust neural correlates of empathy.

However, using GSP for fMRI empathy analysis has limitations, as graph construction varies with the solver used and can be computationally intensive. For future investigations, exploring graph wavelet transform-based analyses can be used for a more in-depth examination of spectral bands, providing deeper localization within the brain. Additionally, a comparative analysis involving resting-state fMRI, a condensed version of the movie, and a structural connectivity network could offer valuable insights. Another research direction involves incorporating prior information from structural connectivity into the learning process for functional connectivity, opening new possibilities for a comprehensive understanding of neural interactions.



\bibliographystyle{IEEEbib} 
\bibliography{references}

\end{document}